# Crystal chemical simulation of superconductors on the basis of oxide and intermetallic layers

L M Volkova

Institute of Chemistry, Far Eastern Branch of the Russian Academy of Sciences
690022 Vladivostok, Russia

E-mail: volkova@ich.dvo.ru

**Abstract**
Simulation of "hybrid" superconductors of 3d-, 4d- and 5d-transition elements consisting of two different superconducting fragments located between positively charged ions planes – B'$O_2$ oxide planes and $B_2C_2$ intermetallic layers – has been performed on the basis of the structure of $Sr_2Mn_3As_2O_2$ ($A_2(B_2C_2)(B'O_2)$). The oxide planes are similar to those of $CuO_2$ in high-temperature superconducting cuprates while the intermetallic layers – to those of $Ni_2B_2$ in low-temperature superconducting borocarbides $RNi_2B_2C$ and $Fe_2As_2$ layers in high-temperature superconducting oxypnictides $RFeAsO_{1-x}F_x$.

## 1. Introduction

The relation between the temperature of transition into superconducting state and the structural parameters in layered superconductors has been proved experimentally. The $T_c$ change can not be always explained by changing of the charge carriers' concentration. The size effect of A cations located in outer planes of the A2CuO$_2$ sandwich on $T_c$ has been known since the discovery of high-$T_c$ cuprates [1, 2]. Moreover, it was shown in [3] that not only the size of A cation, but also that of the doping cations (at fixed hole concentration and lattice parameter) affected $T_c$ as well. The sensitivity of superconductivity to the distances from the CuO$_2$ plane to adjacent A-cations planes [$d$(CuO$_2$-A)] was reported in [4-11], while its sensitivity to the distances between A cations and oxygen atoms from the CuO$_2$ plane [$d$(A-O)] was found in [12]. $T_c$ of the Ln$_{1.9}$Sr$_{0.1}$CuO$_4$ superconductor was doubled (from 25 to 49 K) under effect of strong epitaxial compression, which induced the *a* parameter decrease by 0.02 Å and the *c* parameter increase by 0.1 Å [13]. In addition to the size of A cation and its distance to the CuO$_2$ plane, the charge of this cation also affects $T_c$. $T_c$ is higher in those optimally doped cuprates (with the same number of CuO$_2$ planes) where the charge of A cations A is lower. Variation of $T_c$ along with that of the electric field created by the A cations charges was mentioned in [14]. The studies of diborides MgB$_2$ [15-18] and borocarbides RNi$_2$B$_2$C [19-25] showed that $T_c$ of these superconductors, as in case of HTSC cuprates, depended on structural parameters.

We have previously stated [26-31] the correlations between the superconducting transition temperature ($T_c$) and the relations inside the group of crystallochemical parameters of cationic sublayer for layered superconductors: HTSC on the basis of cuprates, diborides (AB$_2$) and nickel borocarbides (RNi$_2$B$_2$C). These correlations are of similar character for all three classes of compounds, despite the differences in their nature. The correlations enabled us to conclude that an anisotropic three-dimensional fragment – a sandwich with cationic sublattice of $A_2(Cu)$ in HTSC cuprates, $A_2(B_2)$ in diborides and $R(Ni_2)$ in borocarbides – rather than a separate plane was responsible for emerging of the superconducting phenomenon in layered superconductors. The sandwich internal plane contains square networks of Cu atoms in cuprates and Ni atoms in borocarbides or hexagonal networks of boron atoms in diborides, while the external planes are formed by networks of positively charged ions. The critical structural parameters affecting $T_c$ include the distances between Cu, Ni or B atoms in the sandwich internal plane and those characterizing the space between the internal plane and the A cations plane in this sandwich such as: the interval between these planes surfaces, the degree of heterogeneity of the A cations planes surfaces, and the field formed by A and doping cations charges. We showed that the above parameters, aside from their inherent effect on superconductivity, contained information on the charge carriers' concentration. To attain high transition temperatures, the charges of carriers in the sandwich internal plane and of A ions in the external plane must be of the same sign. It is likely that the $T_c$ could also result from overlapping of the space between the sandwich internal and external planes by the coupling electrons (for example, in borocarbides – by the Ni-B covalent bond electrons) [29]. In the superconducting cuprates the coupling electrons between copper ions and axial oxygen atoms are removed from the $CuO_2$ plane due to the Jahn-Teller effect while in $MgB_2$ they are, on the contrary, brought closer to the B2 plane due to substantially ionic character of the Mg-B bond.

The sandwich with a cationic sublattice $R(Fe_2)$ similar to that found in borocarbides $RNi_2B_2C$ (figure 1) can be also discerned in the recently discovered new family of oxypnictides $RFeAsO_{1-x}F_x$ composed of alternating $La_2O_{2-x}F_x$ and $Fe_2As_2$ layers [32-35] with transition temperatures $T_c$ of 25-28 K, which can be raised up to 55 K by replacing La by Nd, Pr and Sm [36-38],

The objective of the present study is to simulate "hybrid" superconductors containing two different superconducting fragments: oxide planes of $CuO_2$-type serving as a basis for high-temperature superconducting cuprates and intermetallic layers of $Ni_2B_2$- and $Fe_2As_2$-type, which participate in formation of superconducting oxypnictides $RFeAsO_{1-x}F_x$ and borocarbides $RNi_2B_2C$, respectively. An additional requirement in regard to simulation consists in placement of both these fragments between positively charged ions planes to form a superconducting sandwich.

## 2. Selection and characterization of structural type

The crystal structure of $Sr_2Mn_3As_2O_2$ belonging to the class of oxidpnictides $A_2Mn_3C_2O_2$, where A= Sr, Ba and C = As, Sb, Bi [39], appears to be appropriate for simulation. Let us designate these compounds as $A_2(B_2C_2)(B'O_2)$. The structure of $A_2Mn_3C_2O_2$ compounds is formed by alternating intermetallic layers $(B_2C_2)$ and metalloxide planes $MnO_2$ $(B'O_2)$. The space between layers is filled by networks of rare earth metal cations. Figure 1 shows the structures formed by $B_2C_2$ and/or $B'O_2$ layers.

The intermetallic $B_2C_2$ layers have been studied as extensively as the metalloxide $CuO_2$ layers. They belong to different structural types such as $ThCr_2Si_2$, $CaBe_2Ge_2$, PbFCl, PbO [40-43] and superconducting phases of $LnNi_2B_2C$ [44] and $RFeAsO_{1-x}F_x$ [32-38] types. The $B_2C_2$-layer can be represented as built from edge-linked tetrahedral $BC_4$ $(CB_4)$ or basal edge-linked square pyramids $CB_4$ $(BC_4)$. Besides, its structure can be considered as a sandwich with inside $B_2$ square networks and outside C networks with square sides equal to $a\sqrt{2}/2$ and $a$ period, respectively. Interchangeability of metal B and nonmetal C positions in the $B_2C_2$-layer provides additional possibilities for variation of not only the value, but also of the sign of the layer charge. This layer is capable to sustain significant distortions and exist not only in a tetragonal system, but also, for example, in a rhombic system under $a$ axis compression like in CuTe ($a$=3,149 Å, $b$=4,086 Å, $c$=6,946 Å) [45]. Moreover, a remarkable feature of this layer is its stability to extensive variations of atoms sizes and oxidation states as well as the valent electrons concentrations.

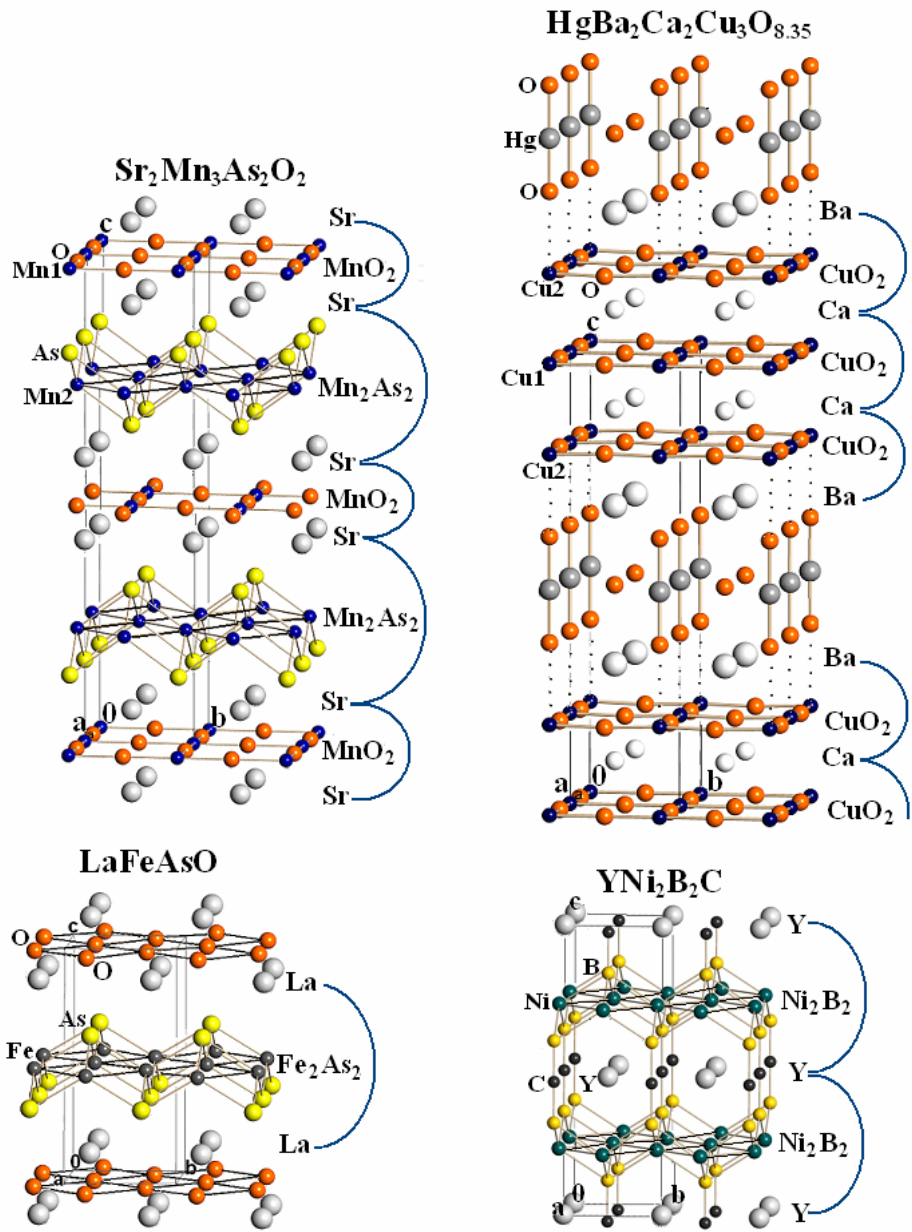

**Figure 1.** The crystal structures of $Sr_2Mn_3As_2O_2$ and superconductors formed by $B_2C_2$ and/or $B'O_2$ layers. The superconducting fragments-sandwiches are shown by semicircles.

The reason of such a stability consists in formation of additional bonds between atoms of the same type (B or C) in flat square networks of both internal and external layers of a $B_2C_2$-sandwich and variation of B-C bond lengths. The following examples demonstrate the layer ability to adapt to the compound composition. The distances Cu-Cu, Si-Si and Cu-Si in the $Cu_2Si_2$-layer are equal to 2.64, 3.73 and 2.52 Å in $HfCuSi_2$ [45] and 2.97, 4.20 and 2.47 Å in $SrCu_2Si_2$ [46], respectively. The distances Te-Te in the $Cu_2Te_2$-layer are equal to 3.149 and 4.086 Å in the compound CuTe [47] and 4.384 Å in the compound NaCuTe [48], respectively. The shortest distances Cu-Cu in $HfCuSi_2$ and Te-Te in CuTe are possibly of a binding character, since they are close to Cu-Cu distances in metallic copper (2.56 Å) and regular Te-Te covalent bond lengths (2.69-2.80 Å) [49]. The $B_2C_2$-layer serves as a buffer to sustain the crystal electroneutrality due to changing the B and C valent states and emerging of the B...B and C...C interactions.

## 3. Candidate superconductor compounds

The number of representatives of the structural type $Sr_2Mn_3As_2O_2$ must be increased with using transition d-elements such as Cu, Nd, Ta, Re, Ru and Ti. The chemical composition $A_2(B_2C_2)(B'O_2)$ of simulated superconductors shall satisfy the following requirements: B' and B (in some cases only B) shall be transition metals in a mixed-valency state with the intervalent charge transfer energy not higher than for copper ions and have a magnetic moment in one of the valent states.

The $B_2C_2$-layers with the structures of $ThCr_2Si_2$- and $CaBe_2Ge_2$-types occur for most of 3d, 4d and 5d elements. Aside from similarity of chemical nature and size, d-elements isomorphism in these layers is facilitated by metal and nonmetal atoms locations in some layer planes.

The question on the possibility of formation of flat $CuO_2$-type layers by these metals remains unanswered. Nevertheless, according to [50], an isomorphic interchangeability of $Cu^{2+}$ by $Mn^{2+/3+}$, $Zn^{2+}$, $Nb^{4+}$, $Re^{4+}$ and $Ta^{5+}$ ions is possible. For the first two Mn ions and, most surprisingly, Zn ions, the compounds $A_2Mn_3Pn_2O_2$ (A = Sr; Pn = P, As, Sb, Bi and A = Ba; Pn = P, As, Sb) [39, 51] and $A_2Zn_3As_2O_2$ (A = Sr, Ba) [52], where Mn and Zn are located in a flat square surrounding by oxygen atoms, have been obtained. Conventionally, a flat coordination is characteristic for Jahn-Teller ions; however, the "forced" coordination phenomenon is known and possibly takes place in $A_2Zn_3As_2O_2$ compounds. It is not also ruled out that coordination of some d-elements is supplemented by two additional oxygen atoms with formation of an octahedron.

There exist some limitations imposed on the $A_2(B_2C_2)(B'O_2)$ compounds composition. In the homologous series of these compounds, just like in the $AB_2C_2$ ($ThCr_2Si_2$- and $CaBe_2Ge_2$-type) compounds series, the value of $a$ parameter varies, as a rule, proportionally to variation of A, B and C atoms sizes. Furthermore, in $A_2(B_2C_2)(B'O_2)$ geometrical parameters of the $B_2C_2$-layer and A ions sizes are additionally affected by the possibility of varying the B'-O bonds lengths in the oxide layer. For example, substitution of Sr by Ba and As by Sb would result in increasing $a$ by approximately 0.1 Å.

Excessive increase of this parameter at substitution of A, B or C by larger ions might result in the B'-O bond rupture and compound destruction. For example, while in the series $A_2Mn_3As_2O_2$ and $A_2Mn_3Sb_2O_2$ the compounds with A = Sr and Ba were obtained, in the series $A_2Mn_3Bi_2O_2$ – only the strontium compounds. This must be also the reason why the authors of [52] could not synthesize the compounds $A_2Zn_3Sb_2O_2$ with antimony along with successful synthesis of $A_2Zn_3As_2O_2$ (A = Sr, Ba) with arsenic.

Let us calculate geometrical parameters of predicted compounds $A_2(B_2C_2)(B'O_2)$. The B'-O bonds lengths in the plane $B'O_2$ for B' = Cu [53], Nb [54 - 56], Ta [57 - 59], Re [60], Ru [61] and Ti [62] would be approximately equal to average bond lengths in $B'O_6$ octahedra of respective elements and fall into the range 1.90-2.00 Å (the limit of stability of B'-O bonds in these elements octahedra is rather wide (1.70-2.30 Å)). Therefore, the elementary unit parameter $a$ and the distances B-B or C-C in the $B_2C_2$-sandwich internal plane would be in the range 3.80-4.00 Å and 2.76-2.83 Å, respectively. The latter values are close to the metal-metal distances found for metals under study (2.64-2.86 Å) [63], except Cu (d(Cu-Cu) = 2.56 Å).

The stability of the $AB'O_2$-fragment in the structure of $A_2(B_2C_2)(B'O_2)$ is determined by the value of tolerance factor ($t$), since this fragment has a perovskite-type structure. When the $a$ unit parameter is equal to 3.9 ±0.1 Å, the most appropriate ions for A position are Sr, Ca and Na, since for them at $t$ in the range of the structure stability (0.8-1) the parameter $a$ varies in the ranges 3.62-4.52, 3.39-4.24 and 3.42-4.28 Å, respectively (ionic radii for calculation are taken from [64]).

Using the above considerations and calculated geometrical parameters, we have performed simulation of previously unknown layered compounds $A_2(B_2C_2)(B'O_2)$ with the structure of $Sr_2Mn_3As_2O_2$-type. Table 1 gives the composition and charge (Q) for the 2A, $B'O_2$ and $B_2C_2$ layers of these systems and the $B_2C_2$-layer characteristics. Table 1 also includes the type of interaction within the plane and distribution of metal (B) and nonmetal (C) atoms on planes (atoms located in the base of $CB_4$ ($BC_4$) pyramid form an internal plane while those located in the top – external planes of the $B_2C_2$-layer).

**Table 1.** Composition and charge (Q) of layers in compounds $A_2(B_2C_2)(B'O_2)$ & $B_2C_2$-layer data

| 2A | $B_2C_2$ | $B'O_2$ | Square pyramids in $B_2C_2$-layer | Interaction type in $B_2C_2$-layer planes |
|---|---|---|---|---|
| $Ca_{2-x}Na_x$ | $B_2C_2$ (B = Mn, Fe, Cu) (C = P, As, Si, Ge) | $CuO_2$ | $CB_4$ | B...B |
| $4-\delta$ | $-2$ | $-2+(+\delta)$ | | |
| $Ca_{2-x}Y(Ln)_x$ | $B_2C_2$ (B = Mn, Fe, Cu) (C = P, As, Si, Ge) | $CuO_2$ | $CB_4$ | B...B |
| $4+\delta$ | $-2$ | $-2+(-\delta)$ | | |
| $Ca_{2-x}Na_x$ | $B_2C_2$ (B = Ru, Ti) (C = P, As) | $B'O_2$ (B' = Ru, Ti) | $CB_4$ | B...B |
| $3-\delta$ | $-2$ | $-1+(+\delta)$ | | |
| | (C = Si, Ge) | (B' = Ru, Ti) | | |
| $4-\delta$ | $-3$ | $-1+(+\delta)$ | | |
| $Na_{2-x}Ca_x$ | $B_2C_2$ (B = Ru, Ti) (C = Si, Ge) | $B'O_2$ (B' = Ru, Ti) | $CB_4$ | B...B |
| $2+\delta$ | $-2$ | $0+(-\delta)$ | | |
| $Na_{2-x}$ | $B_2C_2$ (B = Nb, Ta, Re) (C = Si, Ge) | $B'O_2$ (B' = Nb, Ta, Re) | $CB_4$ | B...B |
| $2-\delta$ | $-2$ | $0+(+\delta)$ | | |
| $Ca_{2-x}Na_x$ | $Cu_2Te_2$ | $CuO_2$ | $TeCu_4$ | Te...Te |
| $4-(\delta_1+\delta_2)$ | $-2+(+\delta_1)$ | $-2+(+\delta_2)$ | | |
| $Y_{2-x}Ca_x$ | $Hg_2Te_2$ | $CuO_2$ | $TeHg_4$ | Hg...Hg Te...Te |
| $5-(\delta_1+\delta_2)$ | $-3+(+\delta_1)$ | $-2+(+\delta_2)$ | | |
| $Ca_{2-x}Na_x$ | $In_2Bi_2$ | $CuO_2$ | $BiIn_4$ | In...In Bi...Bi |
| $4-(\delta_1+\delta_2)$ | $-2+(+\delta_1)$ | $-2+(+\delta_2)$ | | |
| $Na_{2-x}$ | $In_2Bi_2$ | $B'O_2$ (B' = Nb, Ta, Re) | $BiIn_4$ | In...In Bi...Bi |
| $2-(\delta_1+\delta_2)$ | $-2+(+\delta_1)$ | $0+(+\delta_2)$ | | |

Let us present some arguments in favor of such systems realization.

The system $Ca_{2-x}Na_x(B_2C_2)(CuO_2)$ doped by holes ($CuO_2$ layer charge is equal to $2+(+\delta)$) and the system $Ca_{2-x}Y(Ln)_x(B_2C_2)(CuO_2)$ doped by electrons ($CuO_2$ layer charge is equal to $2+(-\delta)$), where B= Mn, Cu, Fe and C = P, As, Si and Ge, are built, according to [50], through substitution of the following elements in $Sr_2Mn_3Pn_2O_2$ (Pn = P, As, Sb, Bi) [39, 52] by isomorphous elements: Sr by Ca(Na) or Ca(Y, Ln); Mn by Cu and Fe; P by Si; Sb by Ge. The $Mn_2Ge_2$ layers were found in $AMn_2Ge_2$ (structural type $ThCr_2Si_2$), where A = Ca, Sr [65].

The possibility of realization of the systems $Ca_{2-x}Na_x(B_2C_2)(B'O_2)$ and $Na_{2-x}Ca_x(B_2C_2)(B'O_2)$, where B and B' = $Ru^{3+}$-$Ru^{4+}$ and $Ti^{3+}$-$Ti^{4+}$ and C = P, As, Si and Ge, is based on isomorphism [50] of the elements Mn, Ru and Ti and the existence of the following intermetallic compounds that are isostructural to $ThCr_2Si_2$: $ARu_2P_2$, $ARu_2As_2$ [66], $ARu_2Ge_2$ [67], where A = Ca, Sr, as well as in $ARu_2Si_2$, and $ARu_2Ge_2$ (A = Y, Ln) [68].

Unfortunately, we could not find data for the system $Na_{2-x}(B_2C_2)(B'O_2)$, where B and B' = Nb, Ta, Re, and C = Si, Ge. However, the elements B and B' included into this system are known [50] to be isomorphous to Cu.

The systems $Ca_{2-x}Na_x(B_2Te_2)(CuO_2)$ and $Y_{2-x}Ca_x(B_2Te_2)(CuO_2)$ are capable to exist with the layers $B_2Te_2$ (B = Cu, Hg), where tellurium atoms are located in external planes while Cu or Hg atoms – in internal planes of this layer. This results from possible isomorphism between Hg, Cu and Fe elements [50], and for the latter two elements such layers were found in the compounds NaCuTe [48], CuTe [47], $FeTe_{0.9}$ and $Fe_{1.125}Te$ [66]. One should mention that the formal valency of mercury in the layer $Hg_2Te_2$ (d(Hg-Hg) ~2.76 Å) would be less than 1, since in $Hg_3AsF_6$ [67], where the distance d(Hg-Hg) = 2.64 Å, the Hg valency is equal to 1/3.

Realization of the systems $Ca_{2-x}Na_x(In_2Bi_2)(CuO_2)$ and $Na_{2-x}(In_2Bi_2)(B'O_2)$, where B' = Nb, Ta, Re, must be also possible. The calculated with using the applied model distances between In atoms (d(In-In)~2.76 Å) located in the internal plane of the $In_2Bi_2$ layer are close to those found, for example, in InSe (d(In-In) = 2.82 Å) [68]. The distances between bismuth atoms (d(Bi-Bi) ~3.9 Å) located in the external planes of the $In_2Bi_2$ layer would be much shorter than the Bi van der Waals diameter (4.4 Å) that enables one to suggest the emerging of Bi...Bi interactions.

## 4. Conclusions

The simulation was performed for new "hybrid" superconductors $A_2(B_2C_2)(B'O_2)$ consisting of two different superconducting fragments: oxide planes $B'O_2$ and intermetallic layers $B_2C_2$ located between positively charged ions planes. The crystal structure of $Sr_2Mn_3As_2O_2$ ($Sr_2(Mn_2As_2)(MnO_2)$) related to one of the crystal chemical factors participating in creating conditions for emerging of superconducting state in layered superconductors was used in the simulation. This structure consists of fragments comprising sandwiches *I/S/I*, in which the internal layer *S* ($B_2C_2$ and $B'O_2$) is capable to accumulate transferrable charge carriers at doping other layers in the structure while the external layers *I* are dielectric planes built from positively charged ions. The possibilities of realization of the $A_2(B_2C_2)(B'O_2)$ system on the basis of B and B' = Cu, Mn, Fe, Nb, Ta, Re, Ru or Ti and C = Si, Ge, P, As, Bi or Te are discussed.

## References

[1] Cava R J 1998 *Nature* **394** 127
[2] Williams G V M and Tallon J L 1996 *Physica C* **258** 41
[3] Attfield J P, Kharlanov A L and McAllister J A 1998 *Nature* **394** 157
[4] Zhang X, Lu W H and Ong C. K 1997 *Physica C* **289** 99
[5] Volkova L M, Polishchuk S A, Magarill S A and Borisov S V 1989 *Sverkhprovodimost: Fiz., Khim., Tekh.* **2** 127
[6] Morosin B, Venturing E L, Dunn R G and Newcomer P P 1997 *Physica C* **288** 255
[7] Nobumasa H, Shimizu K and Kawa T 1990 *Physica C* **167** 515
[8] Volkova L M, Polishchuk S A, Magarill S A and Borisov S V 1991 *Sverkhprovodimost: Fiz., Khim., Tekh.* **4** 155
[9] Motida K 1991 *J. Phys. Soc. Jpn.* **60** 3194
[10] Volkova L M, Polishchuk S A, Magarill S A and Buznik V M 1997 *Dokl. Ross. Akad. Nauk* **353** 200
[11] Varela M, Arias D, Sefrioui Z, Leon C, Ballesteros C, Pennycook S J and Santamaria J 2002 *Phys. Rev. B* 66, 134517
[12] Oshima A and Motida K 2002 *J. Supercond,* **15** 165
[13] Locquet J-P, Ferret J, Fompeyrine J, Maächler E, Seo J W and Van Tendeloo G 1998 *Nature* **394** 453
[14] Amelin I I 1994 *Sverkhprovodimost: Fiz., Khim., Tekh.* **7** 788
[15] Buzea C and Yamashita T 2001 *Supercond. Sci and Techn,* **14** R115-R146
[16] Wan X, Dong J, Weng H and Xing D Y 2002 *Phys. Rev. B* **65** 012502
[17] Neaton J B and Perali A 2001 *Preprint* http://arxiv.org/abs/cond-mat/0104098
[18] Medvedeva N I, Ivanovskii A L, Medvedeva J E and A J Freeman 2001 *Physical Review B* **64** 20502 (Rapid)
[19] Tomy C V, Martin J M and Paul D Mc K 1996 *Physica B* **223&224** 116
[20] Freudenberger J, Kreyssig A, Ritter C, Nenkov K, Drechsler S.-L, Fuchs G, Muller K-H, Loewenhaupt M and Schultz L 1999 *Physica C* **315** 91
[21] Cho B K, Canfield P C and Jonston D C 1996 *Phys. Rev.Lett.* **77** 163
[22] Narozhnyi V N, Fuchs G, Freudenberger J, Nenkov K and Müller K-H 2001 *Physica C* **364-365,** 571
[23] Bitterlich H, Löser W, Behr G, Nenkov K, Fuchs G, Belger A and Schultz L 1998 *Physica C* **308** 243
[24] Jaencke-Roessler U, Belger A, Zahn G, Wehner B, Paufler P and Bitterlich H 1999 *Physica C* **314** 43
[25] Eversmann K, Handstein A, Fuchs G, Cao L and Müller K-H 1996 *Physica C* **266** 27
[26] Volkova L M, Polyshchuk S A, Magarill S A and Sobolev A N 2000 *Inorganic Materials* **36** 919
[27] Volkova L M, Polyshchuk S A and Herbeck F E 2000 *J. Supercond.* **13** 583
[28] Volkova L M, Polyshchuk S A, Magarill S A and Herbeck F E 2001 *J. Supercond.* **14** 693 (*Preprint* http://arxiv.org/abs/cond-mat/0108295)
[29] Volkova L M, Polyshchuk S A, Magarill S A and Herbeck F E 2002 *J. Supercond.* **15** 667 (*Preprint* http://arxiv.org/abs/cond-mat/0205533)
[30] Volkova L M, Polyshchuk S A, Magarill S A and Herbeck F E 2003 *J. Supercond.* **16** 939 (*Preprint* http://arxiv.org/abs/cond-mat/0212625)
[31] Volkova L M, Polyshchuk S A and Herbeck F E 2004 *Focus on Superconductivity*, ed Barry P. Martins (N.Y.: Nova Science Publisher Inc.) p.83
(*Preprints* http://arxiv.org/abs/cond-mat/0402272 and cond-mat/0310511)
[32] Kamihara Y, Watanabe T, Hirano M and Hosono H 2008 *J. Am. Chem. Soc*. **130** 3296
[33] Zhu X, Yang H, Fang L, Mu G and Wen H-H 2008 *Preprint* http://arxiv.org/abs/0803.1288v1

[34] Chen G F, Li Z, Li G, Zhou J, Wu D, Dong J, Hu W Z, Zheng P, Chen Z J, Luo J L and Wang N L 2008 *Preprint* http://arxiv.org/abs/0803.0128v1
[35] Sefat A S, McGuire M A, Sales B C, Jin R, Hove J Y and Mandrus D 2008 *Preprint* http://arxiv.org/abs/0803.2528
[36] Ren Z-A, Yang J, Lu W, Yi W, Shen X-L, Li Z-C, Che G-C, Dong X-L, Sun L-L, Zhou F and Zhao Z-X 2008 *Preprint* http://arxiv.org/abs/0803.4234)
[37] Ren Z-A, Yang J, Lu W, Yi W, Che G-C, Dong X-L, Sun L-L and Zhao Z-X. 2008 *Preprintt* http://arxiv.org/abs/0803.4283
[38] Ren Z-A, Lu W, Yang J, Yi W, Shen X-L, Li Z-C, Che G-C, Dong X-L, Sun L-L, Zhou F and Zhao Z-X 2008 *Preprint* http://arxiv.org/abs/0804.2053
[39] Brechtel E, Cordier G and Schafer H 1979 *Z. Naturforsch. B* **34** 777
[40] Garnier P, Moreau J and Gavarri J R 1990 *Mater. Res. Bull* **25** 979
[41] Binnie W P 1956 *Acta Crystallogr.* **9** 686
[42] Zang H, Wang Y Y, Zhang H, Dravid V P, Marks L D, Han P D, Payne D A, Radaelli P G and Jorgensen J D 1994 *Nature* **370** 352
[43] Ban Z and Sikirica M 1965 *Acta Crystallogr.* **18** 594
[44] Siegrist T, Zandbergen H W, Cava R J, Krajewski J J and Peck Ir W F 1994 *Nature* **367** 254
[45] Andrukhin L S, Lysenko L A, Yarmolyuk Ja P and Gladyshevskii E I 1975 *Dopovidi. Acad. Nauk Ukr.RSR A* **7** 645
[46] Eisenmann B, May N, Müller W, Schafer H, Weiss A, Winter J and Ziegleder G Z 1970 *Naturforsch. B* **25** 1350
[47] Baranova R V and Pinsker Z G 1973 *Z. Kristallografiya* **18** 1169
[48] Sabelsberg G and Schafer H Z. 1978 *Naturforsch. B* **33** 370
[49] Burns R C, Gillespie W C and Slim D R 1979 *Inorg. Chem.* **18** 3086
[50] Makarov E S 1973 *Atom Isomorphy in Crystals* (Moscow: Atomizdat)
[51] Stetson N T and Kauzlarich S 1991 *Inorg. Chem,* **30** 3969
[52] Brock S L and Kauzlarich S M 1994 *Inorg. Chem.* **33** 2491
[53] Sugij N, Yamauchi H and Izumi M 1994 *Phys. Rev. B* **50** 9503
[54] Andersson S, and Galy J 1969 *Acta Crystallogr. Sect. B* **25** 847
[55] Schtckel K und Müller-Buschbaum H 1986 *Rev. Chimie minerale* **23** 154
[56] Wichmann R und Müller-Buschbaum H 1986 *Z. Anorg. Allg. Chem.* **539** 203
[57] Ishizawa N, Marumo F, Kawamura T and Kimura M 1976 *Acta Crystallogr. Sect. B* **32** 2564
[58] Shannon J and Katz L 1970 *Acta Crystallogr.* **26** 102
[59] Marayama N, Sudo E, Kahi K, Tsuzuki A, Kawakami S, Awano M and Torit Y 1988 *Jap. J. Appl. Phys.* **27** L 1623
[60] Calvo C, NG H N and Chamberland B L 1978 *Inorg. Chem.* **17** 699
[61] Maeno Y, Hashimoto H, Yoshida K, Nishizaki S, Fujita T, Bednorz J G and Lichtenberg F 1994 *Nature* **372** 532
[62] Ruddlesden S N and Popper P 1957 *Acta Crystallogr.* **10** 538
[63] Wyckoff R W G 1963 *Crystal Structures* vol 1 (New-York London Sydney: John Wiley & Sons)
[64] Shannon R D and Prewitt C T 1969 *Acta Crystallogr.* **25** 925
[65] Dorrscheidt W, Neiss N and Schafer H 1976 *Z. Naturforsch. B* **31** 890
[66] Wenski G and Newis A Z. 1986 *Naturforsch. B* **41** 38
[67] Schultz J, Williams J W, Miro N D, MacDiarmid A G and Heeger A J 1978 *Inorg. Chem.* **17** 646
[68] Likforman A, Carré D, Etienne J J and Bachet B 1975 *Acta Crystallogr. B* **31** 1252